\documentclass[pre]{revtex4}
\usepackage{amsmath}
\usepackage{graphicx}
\begin{document}
\title{A Simple Model of Cell Proliferation of Bacteria Using  Min Oscillation}
\author{Hidetsugu Sakaguchi and Yuka Kawasaki}
\affiliation{Department of Applied Science for Electronics and Materials,
Interdisciplinary Graduate School of Engineering Sciences, Kyushu
University, Kasuga, Fukuoka 816-8580, Japan}
\begin{abstract}
A mathematical model of Min oscillation in Escherichia coli is numerically studied. The oscillatory state and hysteretic transition are explained with simpler coupled differential equations. Next, we propose a simple model of cell growth and division using the Min oscillation. The cell cycle is not constant but exhibits fluctuation in the deterministic model.  Finally, we perform direct numerical simulation of cell assemblies composed of many cells obeying the simple growth and division model. As the cell number increases with time, the spatial distribution of cell assembly  becomes more circular, although the cells are aligned almost in the $x$-direction.
\end{abstract}
\maketitle
\section{Introduction}
Cell growth and division are important processes of cell proliferation.  
The Escherichia coli (E.~coli) has been intensively studied as one of the simplest model organism.  E.~coli is one of prokaryotic bacteria without a cell nucleus. In the cell division of E.~coli,  a division septum is formed at the mid-zone of the cell. Adler et al. found E.~coli mutants that could not produce a septum at the mid-zone and generated minicells~\cite{Adler}. Later, it was found that the min proteins: MinC, MinD, and MinE play important roles to determine the mid-zone~\cite{Raskin,Hu,Rowland}. The min proteins tend to be localized at cell poles, which suppresses the formation of FtsZ proteins at the poles. As a result, the FtsZ proteins can be formed only in the center, which leads to the formation of the septum at the mid-zone.

The min proteins are not steadily localized at poles, but exhibit reciprocal oscillation between the two poles. The Min oscillation occurs owing to the interaction between the min proteins. The concentration of MinD is low at the center on average. The low concentration region of MinD becomes a position where a division septum is formed. The min proteins diffuse in the cytoplasma and some of them are adsorbed to the cytoplasmic membrane. The transfer dynamics of MinD and MinE between the cytoplasma and membrane is controlled by the densities of MinE and MinD, respectively. 
Several authors studied theoretically the mechanism of the Min oscillation~\cite{Hale, Meinhart}. Howard et al. proposed a one-dimensional reaction-diffusion equation for the densities $\rho_D,\rho_d,\rho_E$, and $\rho_e$ of MinD and MinE in the cytoplasmic membrane and cytoplasma~\cite{Howard}. 

The Min oscillation generates waves in spatially extended media. Chemical waves of Min oscillation have been investigated in vitro. Loose et al. constructed a system of Min oscillation on a lipid membrane, and found spiral waves on the artificial membrane~\cite{Loose}.   
Vecchiarelli et al. found a variety of patterns including burst patterns under slightly different conditions~\cite{Vecciarelli}. 

 We use the deterministic model equation by Howard et al. for a basic model of simplified cell proliferation, although several authors have studied the effect of fluctuations due to the small number of proteins in one cell.~\cite{Rutenberg} In Section 2, we show the Min oscillation in the model equation and its dynamical transition. In Section 3, we study a simple model of cell growth and division based on the Min oscillation. We show the cell cycle fluctuates in time even if the system is deterministic.  In Section 4, we study a simple model of cell proliferation, and found that the form of cell assembly is approximated at an ellipse and the oblateness decreases with time.  
\section{One-dimensional Model of Min Oscillation}
A model equation for the Min oscillation proposed by Howard et al. is expressed as 
\begin{eqnarray}
\frac{\partial \rho_D}{\partial t}&=&-\frac{\sigma_1\rho_D}{1+b_1\rho_e}+\sigma_2\rho_e\rho_d+D_D\frac{\partial^2\rho_D}{\partial x^2},\nonumber\\
\frac{\partial \rho_d}{\partial t}&=&\frac{\sigma_1\rho_D}{1+b_1\rho_e}-\sigma_2\rho_e\rho_d,\nonumber\\
\frac{\partial \rho_E}{\partial t}&=&-\sigma_3\rho_D\rho_E+\frac{\sigma_4\rho_e}{1+b_4\rho_D}+D_ED\frac{\partial^2\rho_E}{\partial x^2},\nonumber\\
\frac{\partial \rho_e}{\partial t}&=&\sigma_3\rho_D\rho_E-\frac{\sigma_4\rho_e}{1+b_4\rho_D},
\end{eqnarray}
where $\rho_D$ and $\rho_d$ express respectively the densities of MinD in the cytoplasma and membrane and  $\rho_E$ and $\rho_e$ are respectively the densities of MinE in the cytoplasma and membrane. $\sigma_i$ ($i=1,2,3,4$) and $b_i$ ($i=1,4$) are parameters to control the adsorption and desorption of the min proteins on the cytoplasmic membrane. The proteins diffuse only in the cytoplasma. The desorption of MinD in the membrane is facilitated and the adsorption of MinD is suppressed by MinE in the membrane. On the other hand, the desorption of MinE in the membrane is suppressed and the adsorption of MinE is facilitated by MinD in the cytoplasma.  In this paper, we will study this equation more in detail and use it for numerical simulations of cell proliferation.

 In the time evolution of Eq.~(1),  $S_D=\int(\rho_D(x)+\rho_d(x))dx$ and $S_E=\int(\rho_E(x)+\rho_e(x))dx$ are conserved. The units of space and time are $\mu$m and s. In this section, we show some numerical results of Eq.~(1) at $D_D=0.28$, $D_E=0.6$, $\sigma_1=20$, $\sigma_2=0.0063$, $\sigma_3=0.04$, $\sigma_4=0.3$, $b_1=0.028$, $b_2=0.027$, $S_D=3000$, and $S_E=170$, which are biologically relevant values used by Howard et al.~\cite{Howard}. Howard already reported several numerical results of the same model equation, and our numerical results are some supplemental ones to their results. The system size $L$ is changed as a control parameter. The no-flux boundary conditions are imposed at $x=0$ and $L$.   

Figure 1(a) shows the time evolution of the density of MinD: $\rho_D(x)+\rho_d(x)$ for $L=2$. A pulse appears near the center and propagates to the right and then another pulse appears near the center and propagates to the left.  The seesaw type density oscillation of MinD occurs between the left and right ends. The density of MinD just at the center is low. Figure 1(b) shows a long-time average of $\rho_D(x)+\rho_d(x)$ at $L=2$, which has a minimum at $x=L/2$. Another protein FtsZ is produced around the minimum point of MinD, and the septum is formed there, which leads to the cell division. Figure 1(c) shows the difference between the maximum and minimum values of the long-time average of $\rho_D(x)+\rho_d(x)$ as a function of $L$. A stationary state is stable for $L\le 1.33$ and jumps to an oscillatory state at $L=1.34$ when $L$ is increased. The oscillatory state jumps to the stationary state at $L=1.18$ when $L$ is decreased.  
There is a weak hysteresis in the transitions. The MinD oscillation works as a signal for the cell division. Although Howard et al. already showed that the Min oscillation disappear at $L<1.2$~\cite{Howard}, the hysteresis was not reported. The disappearance of the Min oscillation in a small system was also reported by other authors.~\cite{Meacci, Fischer}
 \begin{figure}[t]
\begin{center}
\includegraphics[height=4.cm]{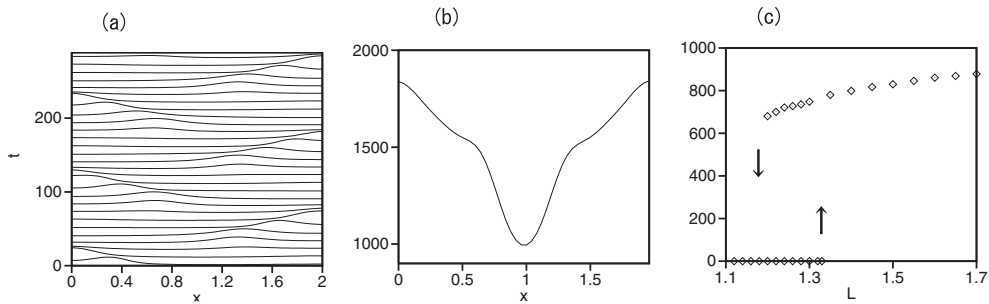}
\end{center}
\caption{(a) Time evolution of the density of MinD: $\rho_D(x)+\rho_d(x)$ for $L=2$. (b) Long-time average of $\rho_D(x)+\rho_d(x)$ at $L=2$. (c) Difference  between the maximum and minimum values of the long-time average of $\rho_D(x)+\rho_d(x)$ as a function of $L$. }
\label{fig1}
\end{figure}

If the densities of MinD and MinE are expanded in Fourier series and only the first two Fourier components are taken, $\rho_D$, $\rho_d$, $\rho_E$, and $\rho_e$ are expressed as 
\begin{eqnarray}
\rho_D(x)&=&X_{D0}+X_{D1}\cos(\pi x/L), \;\rho_d(x)=X_{d0}+X_{d1}\cos(\pi x/L),\nonumber\\ 
\rho_E(x)&=&X_{E0}+X_{E1}\cos(\pi x/L), \;\rho_e(x)=X_{e0}+X_{e1}\cos(\pi x/L).
\end{eqnarray}
The substitution of these expansions into Eq.~(1) yields coupled ordinary differential equations for the Fourier amplitudes $X_{D0}$$\sim$$X_{e1}$:
\begin{eqnarray}
\frac{dX_{D0}}{dt}&=&-\sigma_1 \frac{X_{D0}}{F0}-\sigma_1 \frac{X_{D1}}{b_1 X_{e1}}(1-F_1/F_0)+\sigma_2X_{e0}X_{d0}+0.5\sigma_2X_{e1}X_{d1},\nonumber\\
\frac{dX_{D1}}{dt}&=&-D_D\left (\frac{\pi}{L}\right )^2X_{D1}-2\sigma_1 \frac{X_{D0}}{b_1 X_{e1}}(1-F_1/F_0)-2\sigma_1 \frac{X_{D1}}{(b_1 X_{e1})^2}(F_1^2/F_0-F_1)\nonumber\\
& &+\sigma_2(X_{e0}X_{d1}+X_{e1}X_{d0}),\nonumber\\
\frac{dX_{d0}}{dt}&=&\sigma_1\frac{X_{D0}}{F_0}+\sigma_1\frac{X_{D1}}{b_1X_{e1}}(1-F_1/F_0)-\sigma_2X_{e0}X_{d0}-0.5\sigma_2X_{e1}X_{d1},\nonumber\\
\frac{dX_{d1}}{dt}&=&2\sigma_1\frac{X_{D0}}{b_1X_{e1}}(1-F_1/F_0)+2\sigma_1\frac{X_{D1}}{(b_1X_{e1})^2}(F_1^2/F_0-F_1)-\sigma_2(X_{e0}X_{d1}+X_{e1}X_{d0}),\nonumber\\
\frac{dX_{E0}}{dt}&=&\sigma_4\frac{X_{e0}}{G_0}+\sigma_4\frac{X_{e1}}{b_4X_{D1}}(1-G_1/G_0)-\sigma_3(X_{D0}X_{E0}+0.5X_{D1}X_{E1}),\nonumber\\
\frac{dX_{E1}}{dt}&=&-D_E\left (\frac{\pi}{L}\right )^2X_{E1}+2\sigma_4\frac{X_{e0}}{b_4X_{D1}}(1-G_1/G_0)\nonumber\\
& &+2\sigma_4\frac{X_{e1}}{(b_4X_{D1})^2}(G_1^2/G_0-G_1)-\sigma_3(X_{D0}X_{E1}+X_{D1}X_{E0}),\nonumber\\
\frac{dX_{e0}}{dt}&=&-\sigma_4\frac{X_{e0}}{G_0}-\sigma_4\frac{X_{e1}}{b_4X_{D1}}(1-G_1/G_0)+\sigma_3(X_{D0}X_{E0}+0.5X_{D1}X_{E1}),\\
\frac{dX_{e1}}{dt}&=&-2\sigma_4\frac{X_{e0}}{b_4X_{D1}}(1-G_1/G_0)-2\sigma_4\frac{X_{e1}}{(b_4X_{D1})^2}(G_1^2/G_0-G_1)+\sigma_3(X_{D0}X_{E1}+X_{D1}X_{E0}),\nonumber
\end{eqnarray}
where $F_0=\sqrt{(1+b_1X_{e0})^2-(b_1X_{e1})^2}$, $F_1=1+b_1X_{e0}$, $G_0=\sqrt{(1+b_4X_{D0})^2-(b_4X_{D1})^2}$, $G_1=1+b_4X_{D0}$.
\begin{figure}[t]
\begin{center}
\includegraphics[height=4.cm]{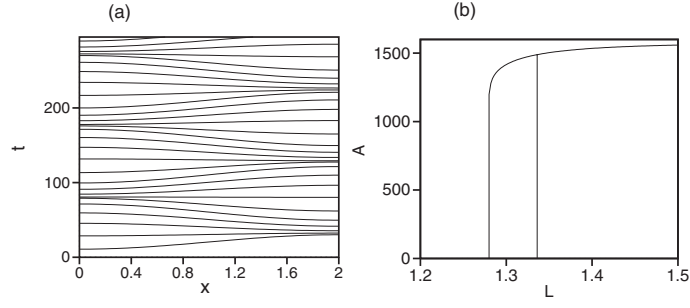}
\end{center}
\caption{(a) Time evolution of the density of MinD: $\rho_D(x)+\rho_d(x)=X_{D0}+X_{d0}+(X_{D1}+X_{d1})\cos(\pi x/L)$ for $L=2$. (b) Peak amplitude of the temporal oscillation of $X_{D1}(t)+X_{d1}(t)$ as a function of $L$.}
\label{fig2}
\end{figure}
\begin{figure}[t]
\begin{center}
\includegraphics[height=4.cm]{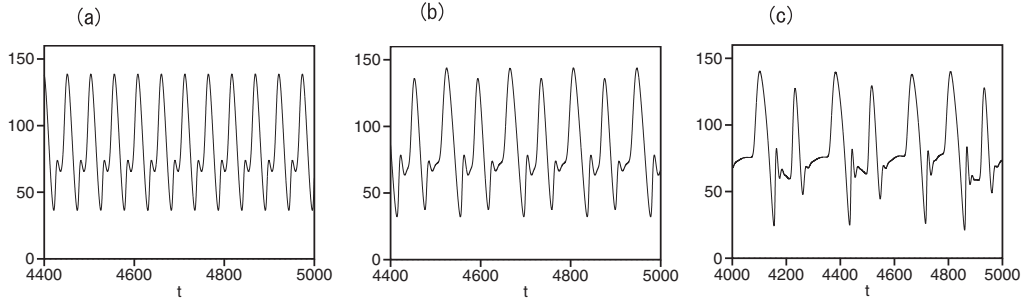}
\end{center}
\caption{Time evolutions of the spatial average of $\rho_D$ 
for (a) $L=2$, (b) $L=2.4$, and (c) $L=3.6$.}
\label{fig3}
\end{figure}
\begin{figure}[t]
\begin{center}
\includegraphics[height=4.cm]{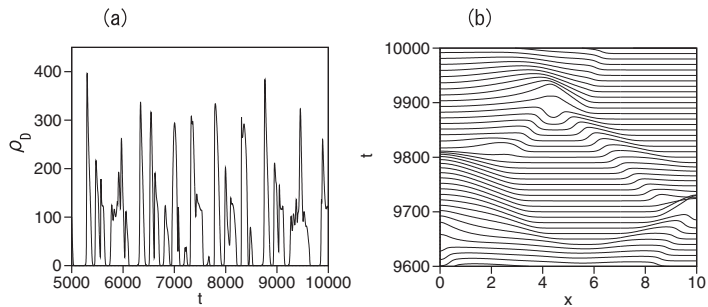}
\end{center}
\caption{Time evolutions of (a) $\rho_D(L/2)$ and profile (b) $\rho_D(x)$
for $L=10$.}
\label{fig4}
\end{figure}

Figure 2(a) shows the time evolution of $X_{D0}+X_{D1}\cos(\pi x/L)+X_{d0}+X_{d1}\cos(\pi x/L)$ at $L=2$. The other parameters are the same as the case of Fig.~1. The seesaw type MinD oscillation appears, however, the pulse propagation is not observed because of the two-mode approximation. Since the long-time average of $X_{D1}$ and $X_{d1}$ is zero, the long-time average of the MinD density is uniform in contrast to Fig.~1(b). Figure 2(b) shows the peak amplitude of $X_{D1}(t)+X_{d1}(t)$ as a function of $L$. The MinD oscillation sets in at $L=1.336$ when $L$ is increased, and the oscillation disappears at $L=1.28$ when $L$ is decreased.  The critical values are slightly different from those for the partial differential equation Eq.~(1), however, similar hysteresis is observed. The two-mode approximation is a useful model to understand the MinD oscillation. 

The Min oscillation becomes more complicated as $L$ is increased. Figure 3 shows the time evolution of the spatial average $\int_0^L\rho_D(x)dx/L$ of $\rho_D$ 
for (a) $L=2$, (b) $L=2.4$, and (c) $L=3.6$. A regular oscillation is observed at $L=2$. The period doubling occurs slightly below $L=2.3$. An oscillation with a double period is observed at $L=2.4$.  An instability of the doubly periodic oscillation occurs at $L\sim 3.5$. More complicated time evolution is observed at $L=3.6$. When $L$ is further increased up to $L=6$, multiple pulses appear. Howard et al. reported that a two-pulse state is stable around $L=8.4$.~\cite{Howard} 
 In such a long cell, the normal cell division by septum formation at the midpoint cannot occur. Figures 4(a) shows the time evolution of $\rho_D(L/2)$ and Fig.~4(b) shows the time evolution of the profile $\rho_D(x)$ at $L=10$. Spatio-temporal chaos with multiple pulses appears in a large system of $L=10$. Wu et al. studied experimentally patterns of the Min oscillation in diverse shapes such as squares, rectangles, circles, and triangle of various sizes.~\cite{Wu} They found the Min oscillation of multiple pulses, and transitions between the two-dimensional patterns with different wavenumbers.  
\section{Simple Model of Growth and Division}
In this section, we study a simple model of cell growth and division based on the equation of the Min oscillation. The cell size is assumed to obey a simple linear growth law $L(t)=L(0)+\Gamma t$. The numerical simulation of Eq.~(1) is performed using the Runge-Kutta method by discretizing the space with an interval  $\Delta x=L/N$ where $N$ is the grid number. In the numerical simulation, the grid interval $\Delta x(t)$ is assumed to increase as $\Delta x(t)=\Delta x(0)+\gamma t$ where $\gamma=\Gamma/N$. The cell division is assumed to occur at the minimum point of the density of MinD.  That is, we assume a simple rule that the cell division occurs at a point where the maximum of the accumulated value $I(x)=\int_0^t(1500-\rho_D(x)-\rho_d(x))dt$ exceeds a critical value $R_c$. We assume that the profiles of $\rho_D$, $\rho_d$, $\rho_E$, and $\rho_e$ are maintained at the cell division. The total grid point are doubled to $2N$ by inserting new grid points at midpoints between neighboring old grid points, and the cell is split into two at a point where $I(x)$ exceeds $R_c$ first. If the cell division occurs at the midpoint, two cells of size $L(t_c)/2$ with the same grid number $N$ are created. However, the profiles of $\rho_D(x)$ et al. in the divided two cells are different, because the profile  of $\rho_D(x)$  before the cell division is neither mirror-symmetric around the midpoint nor spatially-periodic with period $L/2$, that is, two cells of different $\rho_D(x)$ are created at the cell division. This type of numerical simulation of cell division based on the Min oscillation is not performed before.

Firstly, we performed numerical simulations of the simple growth and division model by removing the latter half part of the divided cell. That is, the growth and division of only the former half is repeatedly simulated. Figure 5(a) shows the cell size $L(t)$ at $D_D=0.28$, $D_E=0.6$, $\sigma_1=20$, $\sigma_2=0.0063$, $\sigma_3=0.04$, $\sigma_4=0.3$, $b_1=0.028$, $b_2=0.027$, $S_D=3000$, $S_E=170$, $\gamma=10^{-5}$, and $R_c=1.3\cdot 10^6$. The initial cell size is $L(0)=1.5$ and the grid number is $N=50$. Some spatial inhomogeneity is assumed in the initial condition. The cell size $L(t)$ shows a rather regular oscillation. $L(t)$ grows from $L(t)\simeq 1.4$ and the cell splits into two at $L(t)\simeq 2.8$. 
Since $L(t)$ is always larger that the critical point of $L=1.34$ for the Min oscillation, the Min oscillation is maintained during the growth process. The integral $I(x)=\int(1500-\rho_D(x)-\rho_d(x))dt$ at the midpoint increases steadily and the cell splitting occurs when $I$ goes over the threshold $R_c$.  The profiles of $\rho_D$, $\rho_d$, $\rho_E$, and $\rho_e$ are maintained, however, they are rescaled at the splitting as $\int(\rho_D+\rho_d)dx=S_D$ and $\int(\rho_E+\rho_e)dx=S_E$ are conserved. Figure 5(b) shows the time evolution of cell size $L(t)$ at $R_c=10^6$. The cell size changes randomly between 1.1 and 2.6. For $L(t)<1.18$, the oscillatory state changes into a spatially uniform state and the spatial inhomogeneity decays in time. The dynamics of cell size becomes complicated owing to this transition. 
That is, if $L(t)>1.18$ just after the splitting, $I(x)$ at $x=L/2$ increases steadily and exceeds $R_c=10^6$ at $t=t_c$. In this case, the cell size just before the splitting is relatively small. The cell division occurs at the midpoint and the cell size is reduced to $L(t_c)/2$.  If the cell size $L(t_c)/2$ is smaller than the critical value 1.18 after the splitting, the spatial inhomogeneity decays and $I(L/2)$ increases very slowly, therefore, it takes a large time for $I(L/2)$ to attain the threshold $R_c$.  Then, the cell size just before the splitting is relatively large. Thus, the cell cycle is not constant but changes randomly between 1900 and 3400. That is, the cell cycle exhibits a complex dynamics in the upper level system of growth and division, even if min proteins exhibit a stationary state or a regular limit-cycle oscillation.     

Figure 6 shows the probability distributions of cell cycle for (a) $R_c=1.3\cdot 10^6$, (b) $1.15\cdot 10^6$, and (c) $10^6$. The width of the cell cycle distribution is narrow at $R_c=1.3\cdot 10^6$ because the oscillation of cell size is almost periodic. The width of the distribution increases as $R_c$ decreases. 
For $R_c\le 0.7\cdot 10^6$, $I(x)$ takes a maximum at a point different from the midpoint and an inhomogeneous splitting occurs.  
Figure 7(a) shows the time evolutions of $L(t)$ and $N(t)$ at $R_c=0.7\cdot 10^6$. $L(t)$ changes between 0.92 and 2.86. When $L(t)$ takes a value near $L(t)=2.86$, the inhomogeneous splitting occurs, and the grid number $N$ changes as shown in Fig.~7(b).  
\begin{figure}[t]
\begin{center}
\includegraphics[height=4.cm]{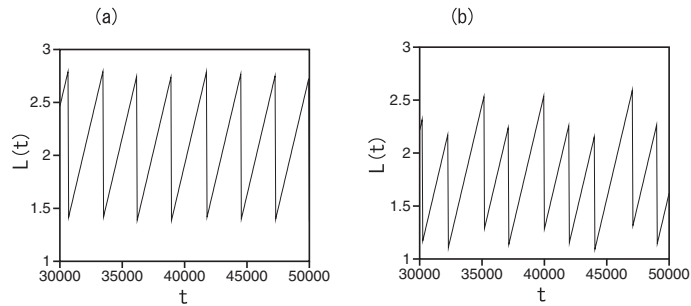}
\end{center}
\caption{Time evolutions of cell size at (a) $R_c=1.3\cdot 10^6$ and (b) $R_c=10^6$ for $L(t)$ at $D_D=0.28$, $D_E=0.6$, $\sigma_1=20$, $\sigma_2=0.0063$, $\sigma_3=0.04$, $\sigma_4=0.3$, $b_1=0.028$, $b_2=0.027$, $S_D=3000$, $S_E=170$, and $\gamma=10^{-5}$. }
\label{fig5}
\end{figure}
\begin{figure}[t]
\begin{center}
\includegraphics[height=4.cm]{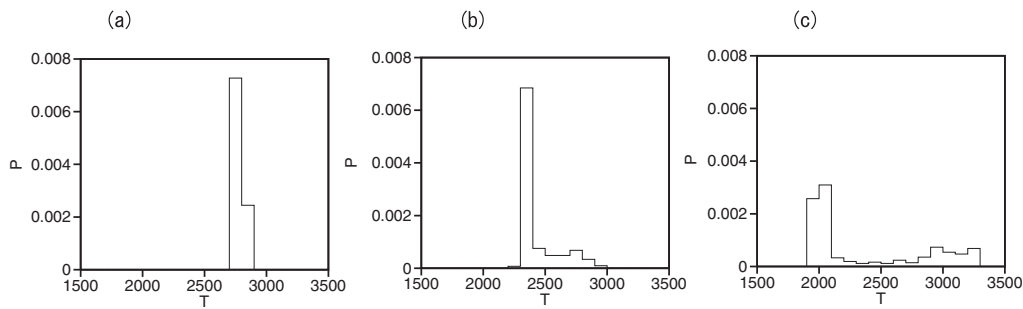}
\end{center}
\caption{Probability distributions of cell cycle at (a) $R_c=1.3\cdot 10^6$, (b) $R_c=1.15\cdot 10^6$, and (c) $R_c=10^6$ for $L(t)$ at $D_D=0.28$, $D_E=0.6$, $\sigma_1=20$, $\sigma_2=0.0063$, $\sigma_3=0.04$, $\sigma_4=0.3$, $b_1=0.028$, $b_2=0.027$, $S_D=3000$, $S_E=170$, and $\gamma=10^{-5}$. }
\label{fig6}
\end{figure}
\begin{figure}[t]
\begin{center}
\includegraphics[height=4.cm]{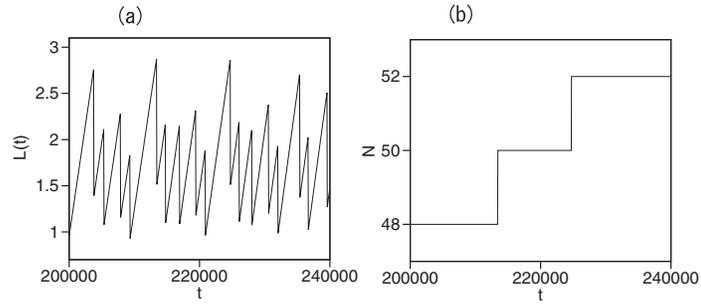}
\end{center}
\caption{Time evolutions of (a) $L(t)$ and (b) $N$ at $R_c=0.7\cdot 10^6$. The other parameters are the same as in Fig.~5.}
\label{fig7}
\end{figure}
\section{Simple Model of Cell Proliferation}
In this section, we study a simple model of cell proliferation. That is, we perform numerical simulation of a growing cell assembly composed of mutually interacting cells. Each cell is assumed to be a rigid rod composed of $N$ points. 
For the cell growth and division, we use the same simple model as studied in the previous section. Langevin type equations of motion are assumed for the coordinates $X_k$ and $Y_k$ of the center of gravity of the $k$th cell and its angle $\Theta_k$ from the $x$ axis. The coordinate of the $i$th grid point in the $k$th cell is expressed as $(x_{i,k},y_{i,k})$ where $x_{i,k}=X_k+\Delta_k(t)(i-N/2)\cos\Theta_k$, $y_{i,k}=Y_k+\Delta_k(t)(i-N/2)\sin\Theta_k$. The interval $\Delta_k$ between the neighboring grid points increases linearly as $d\Delta_k/dt=\gamma$. The grid number $N$ is fixed to be 50, since $R_c$ is set to be larger than $0.8\cdot 10^6$. Repulsive forces are assumed to work when the distance between the grid points of different cells is smaller than 0.05. Furthermore, white noises of variance $3.75\times 10^{-9}$ are applied. 
That is, $X_k$, $Y_k$, and $\Theta_k$ obey the coupled equations:
\begin{eqnarray}
\frac{dX_k}{dt}&=&\sum_{d_{i,j}\le 0.05}\frac{x_{i,k}-x_{j,l}}{d_{i,j}}c(0.5-10d_{i,j})+\xi_{xk}(t),\nonumber\\
\frac{dY_k}{dt}&=&\sum_{d_{i,j}\le 0.05}\frac{y_{i,k}-y_{j,l}}{d_{i,j}}c(0.5-10d_{i,j})+\xi_{yk}(t),\\
\frac{d\Theta_k}{dt}&=&\sum_{d_{i,j}\le 0.05}g\frac{(x_{i,k}-X_k)(y_{i,k}-y_{j,l})-(y_{i,k}-Y_k)(x_{i,k}-x_{j,l})}{d_{i,j}}(0.5-10d_{i,j})+\xi_{\theta k}(t),\nonumber
\end{eqnarray}
\begin{figure}[h]
\begin{center}
\includegraphics[height=4.cm]{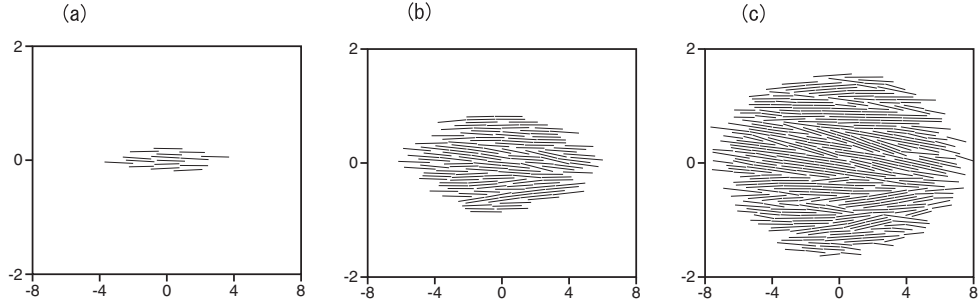}
\end{center}
\caption{Three snapshots of the proliferating cells at (a) $t=10000$, (b) 17500, and (c) 21000. The parameters are $D_D=0.28$, $D_E=0.6$, $\sigma_1=20$, $\sigma_2=0.0063$, $\sigma_3=0.04$, $\sigma_4=0.3$, $b_1=0.028$, $b_2=0.027$, $S_D=3000$, $S_E=170$, $\gamma=10^{-5}$ and $R_c=10^6$. }
\label{fig8}
\end{figure}
\begin{figure}[h]
\begin{center}
\includegraphics[height=4.cm]{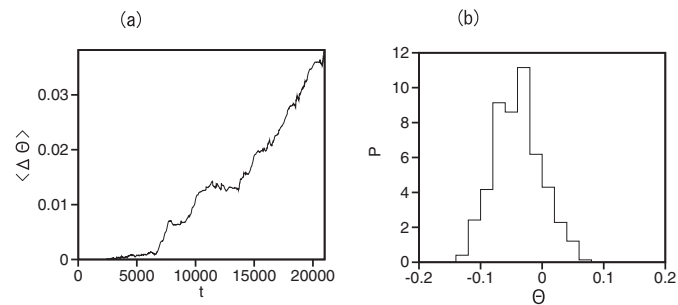}
\end{center}
\caption{(a) Time evolution of $\langle\Theta\rangle$. (b) Probability distribution of $\Theta_i$.}
\label{fig9}
\end{figure}
where $d_{i,j}=\sqrt{x_{i,k}-x_{j,l})^2+(y_{i,k}-y_{j,l})^2}$ is the distance between the two grid points composed of the $k$th cell and the $l$th cell,  and the summation is taken only for the pairs satisfying $d_{i,j}<0.05$. The parameter $c$ is set to be 0.01. 
The third equation is a Langevin equation for the rotation angle $\Theta_k$, and $g/c$ is another parameter related to the moment of inertia, which is set to be 0.2 in our numerical simulation. For the sake of simplicity, we do not consider various effects such as the anisotropy of mobility owing to the rod structure and the change of inertia owing to the cell growth. 
To our best knowledge, this type of numerical simulation of proliferation of linear cells is not reported before. Molecular dynamics simulation of nematic liquid crystals in thermal equilibrium has been performed by many authors.~\cite{Komolkin} 
Recently, active nematics composed of self-propelled rods is intensively studied.~\cite{Sagues} Long-live giant number fluctuations are an interesting topic in the active nematics.~\cite{Narayan} Because our linear cells are not self-propelled, our system is not active nematics. However, our system is considered to be far from equilibrium, since the cell number increases exponentially in time by cell division. We will show a few numerical results for this growing system.   \begin{figure}[h]
\begin{center}
\includegraphics[height=4.cm]{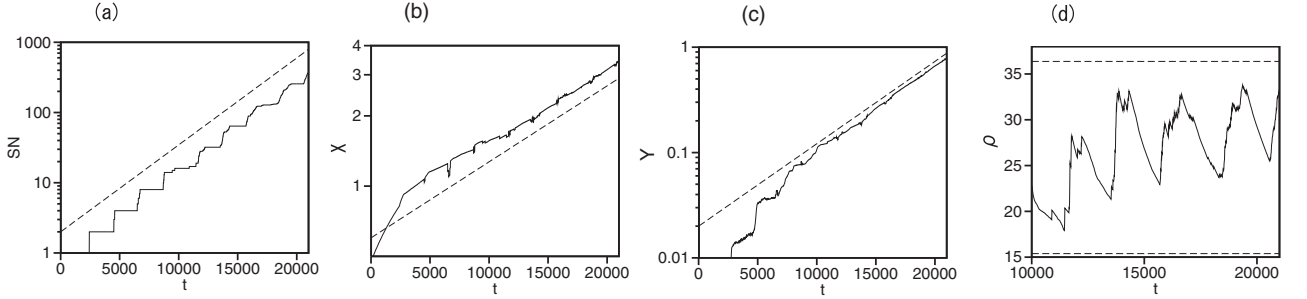}
\end{center}
\caption{Time evolutions of (a) cell number, (b) the first principle component, (c) the second principle component, and (d) the average density $\rho=N/S$ at $R_c=10^6$.}
\label{fig10}
\end{figure}
\begin{figure}[h]
\begin{center}
\includegraphics[height=4.cm]{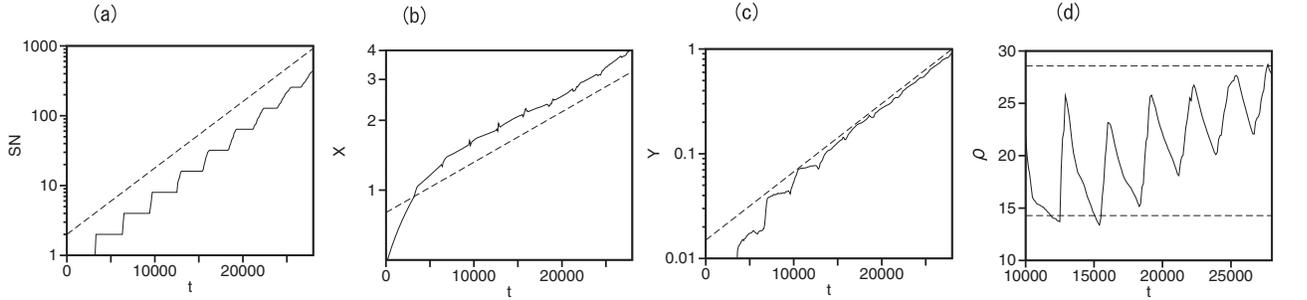}
\end{center}
\caption{Time evolutions of (a) cell number, (b) the first principle component, (c) the second principle component, and (d) the average density $\rho=N/S$ at $R_c=1.3\cdot 10^6$.}
\label{fig11}
\end{figure}
Figure 8(a), (b), (c) are three snapshots of the proliferating cells at $t=10000, 17500$, and 21000. Other parameters are $D_D=0.28$, $D_E=0.6$, $\sigma_1=20$, $\sigma_2=0.0063$, $\sigma_3=0.04$, $\sigma_4=0.3$, $b_1=0.028$, $b_2=0.027$, $S_D=3000$, $S_E=170$, $\gamma=10^{-5}$ and $R_c=10^6$. The cell numbers are respectively 16, 128, and 372 for the three snapshots in Fig.~8.  Roughly speaking, the cell assembly takes an elliptic form. The noise strength is rather small, but the noise term is necessary for the formation of two-dimensional cell assembly. If the noise strength is completely zero, $Y_k=\Theta_k=0$ and cells expand only in the $x$-direction. The cell number increases exponentially and the cell assembly expands only in the $x$-direction as the cells do not overlap with each other, which will be a very cramped growth  

In our model system with small noises, cells can move in the $y$-direction and rotate in the $\theta$ direction. Figure 9(a) shows the time evolution of the standard deviation $\langle\Theta\rangle$ of $\Theta_k$. Figure 9(b) shows the probability distribution of $\Theta_k$ at $t=21000$. The fluctuation of the cell direction increases with time, however, it is still rather small at $t=21000$, that is, all cells are aligned almost in the $x$ direction.  

Figure 10(a) shows the time evolution of the cell number $N$. The cell number increases roughly as $e^{0.000285 t}$. There are two stages of cell division and cell growth. The distribution of the cell assembly is evaluated by the principle component analysis. Figures 10(b) and 10(c) show the time evolution of the first and second principle components denoted by $X$ and $Y$. They are defined as $X=(1/2)\{v_x+v_y+\sqrt{(v_x-v_y)^2+4v_{xy}^2}\}$ and $Y=(1/2)\{v_x+v_y-\sqrt{(v_x-v_y)^2+4v_{xy}^2}\}$ where $v_x=\langle (x_{i,k}-\langle x_{i,k}\rangle)^2\rangle$, $v_y=\langle (y_{i,k}-\langle y_{i,k}\rangle)^2\rangle$, and $v_{xy}=\langle (x_{i,k}-\langle x_{i,k}\rangle)(y_{i,k}-\langle y_{i,k}\rangle)\rangle$. Here, $\langle\cdots\rangle$ denotes the average with respect to $i$ and $k$. Roughly, $X$ and $Y$ increase exponentially as $e^{0.000075t}$ and $e^{0.00018t}$, respectively. The width along the $y$-direction increases more rapidly and the oblateness of the elliptical cell assemby decreases with time. 
The cells grow approximately in the $x$-direction, because they are roughly aligned in the $x$-direction. If the noise term is absent, the cell assembly grows only in the $x$-direction. Therefore, our numerical result that the growth of the cell assembly occurs more rapidly in the $y$-direction is nontrivial. This is a new finding in our numerical simulation, although the mechanism is not sufficiently understood. If the elliptic form is assumed, the area of cells is evaluated as $S=\pi(4/3)XY$. Figure 10(d) shows the time evolution of the average number density $\rho=N/S$. The density increases in the stage of cell division and decreases in the stage of cell growth.  However, the average density tends to increase with time. The dashed lines are $1/(1.1\cdot 0.025)$ and $1/(2.6\cdot 0.025)$. It implies that the interval in the $y$-direction between neighboring two cells is around 0.025, because the cell size changes between 1.1 and 2.6.  It is about half of the interaction range 0.05. Similarly, Fig.~11 shows the time evolutions of (a) the cell number, (b) the first principle component, (c) the second principle component, and (d) the average density $\rho=N/S$ at $R_c=1.3\cdot 10^6$. The cell number increases approximately as $e^{0.000219 t}$. The cell division occurs more synchronously than the case of $R_c=10^6$ shown in Fig.~10(a), because the width of the probability distribution of cell cycle is much narrower at $R_c=1.3\cdot 10^6$. 
 $X$ and $Y$ increase approximately as $e^{0.00005 t}$ and $e^{0.00015 t}$, respectively. The growth rate is slower at $R_c=1.3\cdot 10^6$ than $R_c=10^6$ because the cell cycle is longer at $R_c=1.3\cdot 10^6$. The dashed lines in Fig.~11(d) are $1/(1.4\cdot 0.025)$ and $1/(2.8\cdot 0.025)$. These results show that the cell density exhibits an oscillation and tends to increase with time even at $R_c=1.3\cdot 10^6$. The cell density increases with time probably because it is difficult to keep a constant density by the local repulsive interaction when the cell number expands exponentially.   

\section{Summary}
We have proposed a simple model system of cell proliferation based on the Min oscillation, which describes bacterial dynamics in the wide range from molecular to multi-cellular scales. 

At first, we have shown a few supplemental results for the one-dimensional model of the Min oscillation proposed by Howard et al. We have found a hysteretic transition to the Min oscillation and proposed coupled ordinary differential equations to reproduce the Min oscillation. 

Next, we have performed numerical simulation of a simple model of cell growth and division. The cell cycle does not always take a constant value even in the deterministic model. It is caused by chaotic dynamics of cell size, which is closely related to the transition from the stationary state to oscillatory state. That is, the cell cycle changes with time owing to the interaction of the internal dynamics and the cell growth, even if the system parameters are fixed and no external forces or noises are applied. This phenomenon might be interpreted as an example that different phenotypes can appear even if the genetics and environment are the same. The phenotypic variation in the same genetics and environment has been studied experimentally using E.~coli~\cite{Elowitz}, crayfish~\cite{Vogt}, and so on.    

Finally, we have performed numerical simulation of a simple model of cell proliferation, assuming repulsive interaction between neighboring cells. As the cell number increases, the cell assembly expands in space, and the oblateness of the elliptical cell assembly decreases with time or the cell assembly tends to take a more circular form over time. On the other hand, if noises are completely absent, only one-directional growth occurs in the $x$-direction. We consider that some entropic effect owing to weak noises might cause the more rapid growth in the $y$-direction and decrease of the oblateness. However, our numerical results are preliminary ones and the understanding of the detailed mechanism is left to future study. 

\end{document}